\documentclass[a4paper,12pt]{article}
\usepackage{array}
\usepackage{enumitem}
\usepackage{graphics}
\usepackage{eurosym}
\newcolumntype{L}[1]{>{\raggedright\let\newline\\ \arraybackslash\hspace{0pt}}m{#1}}

\usepackage[left=2cm,right=2cm,top=1.8cm,bottom=1.8cm]{geometry}
\usepackage{graphicx}
\let\OLDthebibliography\thebibliography
\renewcommand\thebibliography[1]{
  \OLDthebibliography{#1}
  \setlength{\parskip}{0pt}
  \setlength{\itemsep}{0pt plus 0.3ex}
}

\usepackage{xspace} % Define commands that appear not to eat spaces

\def\urltilda{\kern -.15em\lower .7ex\hbox{\~{}}\kern .04em}
\def\urldot{\kern -.10em.\kern -.10em}
\def\urlhttp{http\kern -.10em\lower -.1ex\hbox{:}\kern -.12em\lower 0ex\hbox{/}\kern -.18em\lower 0ex\hbox{/}}
\def\sindrumii{{\tt SINDRUM-II}\xspace}
\def\sindrum{{\tt SINDRUM}\xspace}
\def\meg{{\tt MEG}\xspace}
\def\megii{{\tt MEG II}\xspace}
\def\mueii{{\tt Mu2e-II}\xspace}
\def\mgmtwo{{\tt Muon g-2}\xspace}
\def\comet{{\tt COMET}\xspace}
\def\prism{{\tt PRISM}\xspace}
\def\cometi{{\tt COMET Phase-I}\xspace}
\def\cometii{{\tt COMET Phase-II}\xspace}
\def\phasei{{\tt Phase-I}\xspace}
\def\phaseii{{\tt Phase-II}\xspace}

\def\atlas{{\tt ATLAS}\xspace}
\def\cms{{\tt CMS}\xspace}
\def\lhcb{{\tt LHCb}\xspace}
\def\besii{{\tt BES-II}\xspace}
\def\belleii{{\tt Belle-II}\xspace}
\def\jparc{{J-PARC}\xspace}
\def\psi{{PSI}\xspace}
\def\himb{{HiMB}\xspace}
\def\flab{{Fermilab}\xspace}
\def\pipii{{PIP-II}\xspace}
\def\muegamma{\ensuremath{\mu^{+} \rightarrow e^{+}{\gamma}}\xspace}
\def\muec{\ensuremath{\mu^{-}{N} \rightarrow e^{-}{N}}\xspace}
\def\mueee{\ensuremath{\mu^{+} \rightarrow e^{+}e^{-}e^{+}}\xspace}
\newcommand{\mue[1]}{{\tt Mu{#1}e}\xspace}

\begin{document}

\mbox{}
\vskip 5cm
\pagestyle{empty}

\begin{center} \begin{sffamily} \begin{bfseries}
{\Huge Charged Lepton Flavour Violation using Intense Muon Beams at Future 
Facilities}
\end{bfseries} \end{sffamily} \end{center}

\begin{center}
{\large A. Baldini,
  D. Glenzinski,
  F. Kapusta,
  Y. Kuno,
  M. Lancaster, \\
  J. Miller,
  S. Miscetti,
  T. Mori,
  A. Papa,
  A. Sch\"oning,
  Y. Uchida}
\end{center}

\begin{center}
  {\large 
A submission to the 2020 update of the European Strategy for Particle Physics
on behalf of the COMET, MEG, Mu2e and Mu3e collaborations. }
\end{center}

\begin{center} \begin{sffamily} \begin{bfseries}
{\Large Abstract}
\end{bfseries} \end{sffamily} \end{center}

Charged-lepton flavour-violating (cLFV) processes offer deep probes for
new physics with discovery sensitivity to a broad array of new physics
models --- SUSY, Higgs Doublets, Extra Dimensions, and, particularly, models
explaining the neutrino mass hierarchy and the matter-antimatter asymmetry
of the universe via leptogenesis. The most sensitive probes of cLFV
utilize high-intensity muon beams to search for $\mu\rightarrow e$ transitions.

We summarize the status of muon-cLFV experiments currently under
construction at \psi, \flab, and \jparc. These experiments offer
sensitivity to effective new physics mass scales approaching
${\cal{O}}(10^{4})$~$\mathrm{TeV}/c^2$. Further improvements are
possible and next-generation experiments, using upgraded accelerator
facilities at \psi, \flab, and \jparc, could begin data taking within
the next decade. In the case of discoveries at the LHC, they could distinguish among alternative models; even in the absence of direct discoveries, they could establish new physics. These experiments both complement and extend the searches at the LHC.
\vskip 5cm
Contact: Andr\'e Sch\"oning [schoning@physi.uni-heidelberg.de]

\newpage
\pagestyle{plain}
\pagenumbering{arabic}
\setcounter{page}{1}

\begin{sffamily} \begin{bfseries}
{\Large Executive Summary}
\end{bfseries} \end{sffamily}
\begin{itemize}
\item
Charged-lepton flavour-violating (cLFV) processes provide an unique discovery
potential for physics beyond the Standard Model (BSM).
These cLFV processes explore new physics parameter space in a manner
complementary to the collider, dark matter, dark energy, and neutrino physics
programmes.

\item
The global programme includes searches for $\mu\rightarrow e$,
$\tau\rightarrow e$, and $\tau\rightarrow \mu$ transitions at experiments 
hosted in Europe, the US, and Asia. The relative rates among the various 
transitions are model dependent and comparisons among these 
transitions offer powerful model discrimination. A full exploration of cLFV 
parameter space requires the pursuit of all available $\mu\rightarrow e$ and 
$\tau\rightarrow e$,~$\mu$ transitions.

\item
The most sensitive exploration of cLFV is provided by experiments that
utilize high-intensity muon beams to search for cLFV $\mu\rightarrow e$
transitions: a muon decaying into an electron and a photon,
$\mu^{+}\rightarrow e^{+}\gamma$ (\meg\ experiment at \psi); a muon decaying into 
three electrons $\mu^{+}\rightarrow e^{+}e^{-}e^{+}$ (\mue[3] experiment at \psi); and the
coherent neutrinoless conversion of a muon into an electron in the field of a nucleus, \muec\ (\mue[2] experiment at \flab and \comet\ experiment at 
\jparc).

\item
These ``golden'' search channels
%three $\mu\rightarrow e$ transitions
 provide complementary 
sensitivity to new sources of cLFV since the relative rates depend on the 
details of the underlying new physics model.
Thus, it is important to pursue a programme with experiments exploring
all three $\mu\rightarrow e$ cLFV transitions to maximize discovery
potential, and, in the event of discovery, to help differentiate
the various BSM models through a comparison of the rates.

\item
Current limits for cLFV $\mu\rightarrow e$ transitions are in the
$10^{-12}-10^{-13}$ range and probe effective new physics mass scales
above $10^{3}$~$\mathrm{TeV}/c^{2}$. Next-generation experiments
at \meg, \mue[3],
\mue[2], and \comet\ expect to improve these sensitivities by as much as four
orders of magnitude on the timescale of the mid--2020s. 
This
dramatic improvement in sensitivity offers genuine discovery possibilities in a
wide range of new physics models with SUSY, Extra Dimensions, an extended Higgs
sector, lepto-quarks, or those arising from GUT models.

\item
European contributions are vital to the
success of all four of these experiments. Europe hosts two of
them (\meg, \mue[3]) and provides significant detector components for the others
(\mue[2], \comet ).

\item
  Beginning in the latter half of the next decade, upgrades to the beamlines
  at \psi, \flab, and \jparc offer the possibility to further explore 
this parameter space. Improvements in sensitivity by an additional factor 
of 10--100 are possible with: a High intensity Muon Beamline (\himb ) at 
\psi to enable an upgraded \mue[3] (\phaseii); the \pipii linac at \flab 
to enable an upgraded \mue[2] (\mueii); an increased intensity at \jparc 
to enable an upgraded \comet (\phaseii). A next-generation \meg\ 
experiment is also being explored.
Like their predecessors, significant European participation in
the design, construction, data taking, and analysis
will be important to the success of these future endeavors and represents a prudent investment complementary to searches at colliders.

\item
We urge the committee to strongly support the continued participation of
European institutions in experiments searching for cLFV $\mu\rightarrow e$
transitions using high-intensity muon beams at facilities in Europe, the US, and Asia,
including possible upgraded experiments at next-generation 
facilities available in the latter half of the next decade at \psi, \flab, 
and \jparc.
\end{itemize}

\bigskip
\begin{sffamily} \begin{bfseries}
{\Large Objectives}
\end{bfseries} \end{sffamily} 

%%%%%%%%%%%%%%%%%%%%%%%%%%%%%%%%%%%%%%%%%%%%%%%%%%%%%%%%%%%%%%%%%%%%%%%%%%%%%%%%
%
Historically, flavour-changing neutral currents have played a
significant role in revealing details of the underlying symmetries at
the foundation of the SM. In the SM there is no
known global symmetry that conserves lepton flavour.  The discoveries
of quark mixing and neutrino mixing, each awarded Nobel Prizes, provided profound insights to
the underlying physics.  Motivated by these
past successes, there exists a global programme to explore cLFV
processes providing deep, broad probes of BSM physics.

The objective of our programme is to search for evidence of new physics
beyond the SM using cLFV processes in the muon sector. These processes
offer powerful probes of BSM physics and are sensitive to effective
new physics mass scales of $10^3 - 10^4$~$\mathrm{TeV}/c^2$, well
beyond what can be directly probed at colliders. Over the next five
years, currently planned experiments in Europe, the US, and Asia will
begin taking data and will extend the sensitivity to cLFV
interactions by orders of magnitude. The current experiments each
benefit from significant contributions by European institutions. Further improvements
are possible and new or upgraded experiments are being considered that
would utilize upgraded accelerator facilities at \psi,
\flab, and \jparc and could begin taking data in the 2025--2030 timeframe. Strong European  
participation will be important for the success of these next-generation    
muon cLFV experiments.

%%%%%%%%%%%%%%%%%%%%%%%%%%%%%%%%%%%%%%%%%%%%%%%%%%%%%%%%%%%%%%%%%%%%%%%%%%%%%%%%

\bigskip
\begin{sffamily} \begin{bfseries}
  {\Large Scientific Context}
\end{bfseries} \end{sffamily} 

%%%%%%%%%%%%%%%%%%%%%%%%%%%%%%%%%%%%%%%%%%%%%%%%%%%%%%%%%%%%%%%%%%%%%%%%%%%%%%%%

Flavour violation has been observed in quarks and neutrinos, so it is natural to
expect flavour violating effects among the charged leptons as well.  In fact, once
neutrino mass is introduced, the SM provides a mechanism for 
cLFV via lepton mixing in loops. However, the rate is suppressed by factors 
of $\left( \Delta m^{2}_{ij} / M^{2}_{W} \right)^2$, where $\Delta m^{2}_{ij}$ is
the mass-squared difference between the $i^{th}$ and $j^{th}$ neutrino mass
eigenstates, and is estimated to be extremely small, for example
$BF\left( \mu\rightarrow e\gamma \right) \sim 10^{-54}$~\cite{BFmeg-petcov}. 
Many extensions to the standard model predict large cLFV effects 
that could be observed as new experiments begin data taking over the next 
five years. Significant improvements are expected across a wide variety of 
cLFV processes (e.g. $\tau \rightarrow \mu\mu\mu$, $\mu\gamma$, or $e\gamma$; 
$\mu\rightarrow e\gamma$, $eee$; $\mu N\rightarrow e N$; 
$Z$ or $H^{0}\rightarrow e\mu$, $e\tau$, or $\mu\tau$; 
$K_{L}\rightarrow e\mu$). 
The largest improvements are expected in experiments that search for cLFV 
transitions using muons.

Experimentally, there are three primary muon-to-electron transitions used to search for 
cLFV\footnote{Muonium oscillations, $\mu^{+} e^{-}\rightarrow \mu^{-} e^{+}$,
in which the muon and electron form a bound state, can also be used to set limits
on cLFV interactions~\cite{muoniumoscillations} but are not discussed here.}:
a muon decaying into an electron plus a photon, 
$\mu^{+}\rightarrow e^{+}\gamma$; 
a muon decaying into three electrons, \mueee; and direct muon-to-electron 
conversion via an interaction with a nucleus, \muec. These three 
$\mu \rightarrow e$ transitions provide complementary sensitivity to new 
sources of cLFV since the observed rates will depend on the details of the 
underlying new physics model. For example, for models in which cLFV rates 
are dominated by $\gamma$-penguin diagrams, the $\mu\rightarrow e\gamma$ 
transition rate is expected 
to be $\sim 10^2$ times larger than the $\mu\rightarrow eee$ and $\mu{N}\rightarrow 
e{N}$ rates. On the other hand, if the cLFV rates are dominated by $Z$- 
or $H$-penguin diagrams, or if tree level contributions are allowed (e.g. as in 
some lepto-quark or $Z^\prime$ models), then the $\mu\rightarrow e\gamma$ 
rate is suppressed and $\mu\rightarrow eee$ and $\mu{N}\rightarrow e{N}$ 
rates can instead be largest. Thus, a programme with experiments exploring 
all three muon cLFV transitions maximizes the discovery potential and 
offers the possibility of differentiating among various BSM models by 
comparing the rates of the three transitions. This is discussed 
extensively in the literature, see for example 
references~\cite{CalibiSignorelli} and~\cite{Cirigliano}.

Searches for $\mu\rightarrow e$ transitions have been pursued since 1947 
when Pontecorvo first searched for the $\mu\rightarrow e\gamma$ process. Since 
then, the sensitivity has improved by eleven orders of magnitude via a series of
increasingly challenging experiments. The current best limits for the three 
$\mu\rightarrow e$ transitions are 
$BF\left( \muegamma \right) < 4.2\times 10^{-13}$~\cite{MEGLimit}, 
$BF\left( \mueee \right) < 1\times 10^{-12}$~\cite{Mu3eLimit}, 
$R_{\mu e}\left( \mathrm{Au} \right) < 7\times 10^{-13}$~\cite{Mu2eLimit} 
at 90\% CL, where $R_{\mu e}$ is the $\mu\rightarrow e$ conversion rate 
normalized to the rate of ordinary muon nuclear capture. Currently planned 
experiments in Europe, the US, and 
Asia will provide sensitivities well beyond these existing limits. The \meg\ 
experiment at \psi has recently completed an upgrade and expects to extend 
the \muegamma sensitivity by about an order of magnitude with 
%a three year 
physics data taking beginning in 2019. Further improvements will require a new 
approach and/or advances in instrumentation. The first phase of the \mue[3] 
experiment is under construction at \psi and with about 300 days of data 
taking is expected to improve the \mueee sensitivity by over two orders 
of magnitude. A second-phase experiment with additional instrumentation 
could offer a further one order of magnitude improvement with an upgraded 
muon beam providing $>2\times 10^9$~stop-$\mu^{+} / {s}$ (e.g. a 
high-intensity muon beam, \himb, at \psi; or a dedicated $\mu^+$ beamline 
from \pipii at \flab). The \comet\ experiment under construction at \jparc 
will extend the sensitivity to \muec\ by about two orders of magnitude by the 
early-2020s, while the \mue[2] experiment under construction at \flab will 
extend the sensitivity by about four orders of magnitude by the mid-2020s.
%The \mue[2] experiment under construction at \flab 
%and the \comet\ experiment under construction at \jparc will each extend 
%the sensitivity to \muec\ by four orders of magnitude by the mid-2020s. 
On a longer timescale, upgrades in proton intensity offer the possibility of  
additional improvements. An upgrade to \mue[2] that extends the sensitivity 
by another factor of ten or more, \mueii, is proposed and would utilize about 100~kW of 0.8~GeV 
protons from the \flab\ \pipii linac. An upgrade to \comet, \cometii, is proposed and would utilize about 56~kW of 8~GeV 
protons to reach a comparable sensitivity. The status of the currently planned 
experiments and their potential for further improvement is discussed in 
more detail in the next sections.

The outstanding sensitivities that can be achieved by the muon cLFV 
experiments provide access to new physics mass scales in the 
$10^{3} - 10^{4}$~$\mathrm{TeV}/c^2$ range, well beyond what can be 
directly probed at colliders. In general, these experiments explore the BSM 
parameter space in a manner complementary to the rest of the HEP 
experimental programme.

The search for muon-cLFV explicitly probes for flavour-violation
in either CP-conserving or CP-violating BSM interactions; in contrast,
for instance, to muon g-2 which is sensitive to flavour conserving
(and chirality-flipping) interactions. If the {\flab} {\mgmtwo} experiment 
confirms the BNL measurement~\cite{BNL-PRD-2004} and hence an $a_\mu$ value at
odds with the SM beyond 5$\sigma$, it will establish the presence of a
BSM muon interaction which has obvious ramifications for muon-cLFV,
since, in many BSM scenarios, the two are closely related~\cite{gm2_CLFV}. If the $a_\mu$ 
anomaly disappears, the muon-cLFV experiments are still compelling since they
probe effective mass scales well beyond the TeV scale probed by \mgmtwo.

As the charged counterpart of neutrino oscillations,
cLFV plays a significant role in most of the BSM models seeking to explain the
neutrino mass hierarchy and the universe's matter anti-matter
asymmetry generated through leptogenesis.  The cLFV measurements thus
have considerable synergy with the neutrinoless double beta decay and
neutrino oscillation research programmes.  For example, there is a large 
class of models (see e.g.~\cite{SEESAWS}) proposed to explain the 
smallness of the neutrino mass. 
These typically involve extensions to the Higgs sector
and the existence of heavier neutrino partners, the properties of
which --- sterile or non-sterile, Dirac or Majorana, and the mass-scale of
the neutrino partners --- depend on the model. These heavy neutrino
partners typically also play a role in generating a matter anti-matter
asymmetry. The majority of these models predict large cLFV effects, and the
comparison of cLFV and neutrino measurements together becomes a strong
constraint on the model type and its parameters. Indeed, in the most
natural models, where the neutrino partners are extremely massive,
these measurements are one of the few portals into GUT-scale physics.
In the Inverse Seesaw models~\cite{INV-SS}, right-handed neutrinos
with masses in the TeV-scale are produced that are potentially
observable at the LHC. The present LHC limits are below 1~TeV whereas
\mue[2], \comet, and \mue[3] will extend this sensitivity to
2~TeV. More generally \mue[2], {\comet}, and \mue[3] still have a 
sensitivity for
RH neutrinos up to masses of a few PeV, well beyond the direct
detection limit of the LHC. 

The $\mu \rightarrow e$ experiments also provide complementary
information regarding the Majorana nature of neutrinos via the
$\mu^{-}\rightarrow e^{+}$ transition: $\mu^{-} N(Z,A) \rightarrow
e^{+}N(Z-2,A)$. This transition violates both lepton number and lepton
flavour and can only proceed if neutrinos are Majorana.  This search
channel comes for ``free'' in the \mue[2] and {\cometi} experiments.
The \mue[2] and \comet\ sensitivity to Majorana neutrinos will
significantly extend beyond the current best
limit~\cite{MuminusToEplus} with a $\langle{m_{e\mu}}\rangle$ effective Majoarna
neutrino mass scale sensitivity down to the MeV region surpassing the
$\langle{m_{\mu\mu}}\rangle$ sensitivity in the kaon sector which is limited to the
GeV region~\cite{MuminusToEplusProspects}.

%%Lepton number violating processes in muonium-antimuonium oscillations 
%%might also possibly be tested with the 
%%\mue[3] experiment upgraded to include a dedicated low energy positron 
%%detector, similar
%%to~\cite{muoniumoscillations}.

The anomalies in B decays reported by the B-factories and \lhcb and
the {\tt{E821}} $a_\mu$ anomaly have promoted a renewed interest in
leptoquarks~\cite{LQ} and $Z^{\prime}$s~\cite{Z-PRIME}. These models
can generate large cLFV effects via tree-level contributions. Direct
searches for leptoquarks at \atlas and \cms place limits in the
400--800~$\mathrm{GeV}/c^2$ range which will ultimately increase to
approximately 1~$\mathrm{TeV}/c^2$ with HL-LHC. While there are
model-dependencies, the limits from muon cLFV
experiments~\cite{LQ-CLFV} are much stronger with sensitivities up to
masses of 300~$\mathrm{TeV}/c^2$ beyond the present limit
(120~$\mathrm{TeV}/c^2$) established from lepton-flavour violating
B-decays~\cite{LFV-LHCB}.

Many experiments will search for the cLFV $\tau\rightarrow\mu$ and
$\tau\rightarrow e$ transitions
including \atlas, \belleii, \besii, \cms, and \lhcb. In general the
existing limits will be extended by about an order of magnitude to the
$10^{-9} - 10^{-10}$ range. The proposed {\tt {tauFV}}
experiment~\cite{tauFV} may offer another order of magnitude
improvement.  Thus, the ultimate sensitivity offered by the tau-cLFV
searches is several orders of magnitude below the sensitivity offered
by the muon cLFV experiments. The relationship between tau-cLFV and
muon-cLFV processes is model dependent. The large, close-to maximal,
mixing in the neutrino sector favours scenarios in which the rates of
cLFV are similar in the two sectors, but other scenarios are also
possible in which tau-cLFV rates are significantly enhanced. A
comparison of all the transitions: $\mu\rightarrow e$,
$\tau \rightarrow e$ and $\tau \rightarrow \mu$ is a very important
probe of flavour models. All measurements should be pursued.

In summary, experiments sensitive to violations of lepton flavour, lepton number, 
and lepton universality play a significant role in the search for BSM physics. 
It will be necessary to make as broad an array of  measurements as possible in 
order to maximally probe the available parameter space.
The muon-cLFV experiments explore cLFV transition rates that are 
many orders of magnitude beyond what is explored by other experiments and offer
sensitivity to new phenomena with mass scales in the few $\mathrm{PeV}/c^2$ region. 
Over the next several years, the {\meg}, \mue[3], \mue[2], and {\comet} 
experiments have the best reach in their respective channels. Future upgrades 
could extend the sensitivity another one to two orders of magnitude by utilizing 
improved accelerator beamlines, and could begin data taking in the
2025--2030 timescale. These 
future experiments (e.g. \mue[3] \phaseii, \mueii, \cometii, {\tt PRISM}) 
would offer the most sensitive probes of cLFV for the foreseeable future.

%%%%%%

\bigskip
\begin{sffamily} \begin{bfseries}
{\Large Beam Facilities}
\end{bfseries} \end{sffamily}

The muon-cLFV experiments rely on facilities with high-power proton beams
capable of delivering high-intensity muon beams. The experimental infrastructure costs range from \euro\,5--50M with additional substantial facility costs. For example the \mue[2] experiment has a total project cost of  \textdollar\,274M.
Several facilities exist
with proton beams and transport channels
capable of providing muon beams at
high intensities.
The \psi\ laboratory utilizes 1.4~MW of 590~MeV protons to provide
high-intensity beams of secondary particles, including the most
intense low-energy muon beams in the world. The $\rm{\pi E5}$ channel serves
the particle physics community and provides positively charged muons with a
momentum of 28~$\rm{MeV}/c$, a momentum bite of $5-7\%$ FWHM, and
rates up to $10^{8}$~stop-$\mu^{+}/s$. At \flab\ 700~kW of 8.9~GeV/c
protons are available for various experiments, of which about 8~kW will be
utilized to produce about $10^{10}$~stop-$\mu^{-}/s$ for \mue[2].
Similarly, at \jparc\ about 500~kW of up to 30~GeV protons are available
for various experiments, of which about 3~kW (at 8~GeV/c) will be utilized to produce
$10^{9}$~stop-$\mu^{-}/s$ for \cometi.
The \flab\ and \jparc\ muon beamlines are expected to become
operational in the next few years.

Future facilities, capable of providing stopped muon rates a factor of
10--100 larger, are being planned and could become available as early as
2025. These future facilities would enable next-generation muon-cLFV
experiments with improved sensitivities. At \psi, strong requests from
both the particle physics and material
science communities have motivated studies to upgrade the existing muon
beamlines (HiMB study). By optimizing the existing M target station and
improving the transport efficiency,
a new beamline could deliver over $10^{10}$~stop-$\mu^{+}/s$ for \mue[3]
\phaseii\ or a future extension of the \meg\ experiment. At \flab, the
long-baseline neutrino programme motivates the need for a significant
upgrade of the proton beam intensity. The \pipii\ linac will be CW capable
and will use superconducting RF technology to provide 1.6~MW of 0.8~GeV
protons available for a variety of experiments~\cite{PIPII}. Conceptual 
designs exist
to provide about 100~kW of protons to an upgraded \mue[2] experiment,
\mueii, with over $10^{11}$~stop-$\mu^{-}/s$. At \jparc, plans exist to 
provide 56~kW to produce $2\times 10^{11}$~stop-$\mu^{-}/s$ for 
\cometii. Further future, in conjunction with a 1.3~MW
upgrade to the Main Ring at \jparc, the PRISM project would utilize a 
fixed-field alternating gradient (FFAG)
muon storage ring to produce a very intense, very high
purity, monochromatic muon beam with the potential to make a whole programme
of muon-based measurements at world-class sensitivities.

\bigskip
\begin{sffamily} \begin{bfseries}
{\Large Methodology}
\end{bfseries} \end{sffamily}

The same basic experimental methodology is employed in searches for all
three cLFV $\mu\rightarrow e$ processes. The experiment beamline begins
by colliding protons onto a production target to produce low momentum
pions. The resulting pions are either transported through a decay volume or
directly stopped inside the target, and their
decay muons are collected. These experiments require low momentum muons,
typically with momenta less than 50~$\mathrm{MeV}/c$, in order to stop them in
thin targets at the center of the experimental apparatus. At these low momenta,
muons stop in a few~mm or less of material. To reach the target sensitivities
requires high-intensity muon beams, $>10^{8}$~stop-$\mu/s$. The
detector apparatus is designed to precisely determine the energy, momentum, and
timing of particles originating from the muon stopping target. Because these
experiments aim for such extreme sensitivities, their apparatus are customized to the final state of interest.

\medskip
\begin{sffamily}\begin{bfseries}
  {\large The \muegamma\ process}
\end{bfseries}\end{sffamily}
%%%%%%%%%%%%%%%%%%%%%%%%%%%%%%%%%%%%%%%%%%%%%%%%%%%%%%%%%%%%%%%%%%%%%%%%%%%%%%%

The \muegamma\ process is sensitive to new physics mass scales around 
$10^{3}$~$\rm{TeV}/c^{2}$ and primarily tests cLFV dipole couplings where 
new physics appears in loops. The most stringent limit on this process was 
established by the \meg\ experiment using data collected from 2009--2013,
$BF\left( \muegamma \right)<4.2\times 10^{-13}$ at $90\%$ CL~\cite{MEGLimit}.
The \megii\ experiment has recently completed construction and aims to 
improve the sensitivity by an order of magnitude.

The experimental signature of a \muegamma\ decay at rest  
is given by a back-to-back, photon-positron pair coincident in time and 
each with an energy of half the mass of the muon. Each event can 
be described by four observables: the photon and positron energies 
($\rm{E}_\gamma,\, \rm{E}_e$), their relative direction ($\Theta_{e 
\gamma}$), and their relative emission time ($\rm{t}_{e \gamma}$).

The background has two components: an intrinsic physics background coming from
the radiative muon decay (RMD), $\mu \rightarrow e \nu \bar{\nu} 
\gamma$, when the neutrinos carry a small fraction of the available 
energy; and an accidental background that arises when an energetic 
positron from a standard muon decay overlaps with an energetic photon 
from RMD, $e^{+}e^{-}$ annihilation-in-flight, or bremsstrahlung. The
effective branching fraction for the accidental background is a strong 
function of the muon beam intensity, $I_{\mu}$, and the detector resolutions 
associated with the four observables, $\rm{\Delta \rm{E}_{\gamma}}$, 
$\rm{\Delta \rm{E}_{e}}$, $\Delta\Theta_{e\gamma}$, and 
$\Delta \rm{t}_{e\gamma}$:
\begin{equation}
  BF_{\rm{eff}} \propto   
    I_{\mu}^2 \times (\Delta E_{\gamma})^2
\times \Delta E_e \times
    (\Delta\Theta_{e \gamma})^2 \times
\Delta t_{e \gamma}.
\label{Accback}
\end{equation}
In the \meg\ experiment, which used 
$I_\mu\sim 3\times 10^{7}$~stop-$\mu^{+}/s$, the accidental background 
accounted for over $90\%$ of the events near the signal region  
($\rm{E}_{\gamma}>48$~MeV). To achieve an improved sensitivity, \megii\ will 
utilize a higher muon beam intensity. Mitigating the accidental background 
requires \megii\ to upgrade its detector components to achieve the required, improved 
resolutions.
 
%%%%%%%%%%%%%%%%%%%%%%%%%%%%%%%%%%%%%%%%%%%%%%%%%%%%%%%%%%%%%%%%%%%%%%%%%%%%%%%%
\begin{sffamily}\begin{bfseries}
  {\textbf{Status and Plans of the \megii\ experiment}}
\end{bfseries}\end{sffamily}
%%%%%%%%%%%%%%%%%%%%%%%%%%%%%%%%%%%%%%%%%%%%%%%%%%%%%%%%%%%%%%%%%%%%%%%%%%%%%%%%

The \megii\ experiment~\cite{MEGII} is depicted in Fig.~1 of the Addendum. 
The main features are an $e^{+}$ spectrometer formed by a new cylindrical 
drift chamber plus precision pixelated timing counters, located inside a
superconducting solenoid with a graded magnetic field along the beam 
axis, and a $\gamma$ detector, located outside the solenoid, made up of a
homogeneous volume of 900~liters of liquid xenon readout in the central 
region by silicon photomultipliers and in the forward and background 
region by photomultiplier tubes. The finer granularity of the silicon 
photomultipliers provides improved $\gamma$ angular and energy resolution. 
Additional systems are used to further reduce RMD background, and to 
monitor the beam quality and stopping target {\it{in situ}}. The detector
construction is complete and commissioning has begun. Physics data taking 
is expected to begin in 2019 and to last for a few years. The upgraded 
detector is expected to provide resolutions roughly a factor of two better 
than \meg, thus allowing \megii\ to utilize the full muon beam intensity  
available at \psi, $I_{\mu}\sim 10^{8}$~stop-$\mu^{+}/s$, to achieve a factor of ten improvement in expected sensitivity.

%%%%%%%%%%%%%%%%%%%%%%%%%%%%%%%%%%%%%%%%%%%%%%%%%%%%%%%%%%%%%%%%%%%%%%%%%%%%%%%%
\begin{sffamily}\begin{bfseries}
{\textbf{Search for \muegamma\ at future facilities}}
\end{bfseries}\end{sffamily}
%%%%%%%%%%%%%%%%%%%%%%%%%%%%%%%%%%%%%%%%%%%%%%%%%%%%%%%%%%%%%%%%%%%%%%%%%%%%%%%%

Improvements in sensitivity to the \muegamma\ process beyond the \megii\ 
projection may be possible by utilizing the increased muon 
intensities that could be available from future facilities at \psi, \flab, 
and \jparc. However, future experiments would have to devise ways to reduce 
the accidental background below the $10^{-15}$ level in order to fully 
exploit the discovery potential offered by the increased muon intensities.

The use of an active or segmented target could allow a determination of 
the muon decay vertex, which, in principle, should lead to a strong 
suppression of the accidental background. Initial studies~\cite{Papa} made for the \megii\ project
indicated that this additional suppression would be required but this idea could be more effective if different schemes (see below) are adopted.
Improvements to $\Delta\rm{E}_\gamma$ and $\Delta\Theta_{e \gamma}$ 
resolutions should be the most effective given their quadratic effect on 
the accidental background. Feasibility studies have been performed for 
two very different experimental concepts. One is based on a calorimetric 
detection of the photon, like \megii, but with improved energy and timing 
resolutions~\cite{Cavoto}. The other is based on converting the photon and 
precisely measuring the trajectories of the resulting $e^{+}e^{-}$ pair 
with a tracking spectrometer~\cite{Cavoto,Hitlin}. The photon conversion 
concept is also being studied by \mue[3]\ as a potential extension to its 
physics programme. These studies promise sensitivities around $10^{-15}$ at 
$90\%$ CL after a few years running, but additional studies are required to 
verify the efficacy of these concepts.

\medskip
\begin{sffamily}\begin{bfseries}
  {\large The \mueee\ process}
\end{bfseries}\end{sffamily}
%%%%%%%%%%%%%%%%%%%%%%%%%%%%%%%%%%%%%%%%%%%%%%%%%%%%%%%%%%%%%%%%%%%%%%%%%%%%%%%%

The \mueee\ process is sensitive to new physics at mass scales beyond
$10^{3}$~$\rm{TeV}/c^2$ and probes cLFV couplings that arise from 
dipole interactions where new physics appears in loop diagrams and from
$\mu eee$ contact interactions. The most stringent limit on this process 
was established by the \sindrum\ experiment using data collected from 
1983--1986,
$BF\left( \mueee \right) < 1\times 10^{-12}$ at $90\%$~CL~\cite{Mu3eLimit}.
Any improvement in the sensitivity beyond this has a significant impact on 
models predicting cLFV, especially the associated four-fermion couplings.
The \mue[3] collaboration aims for a sensitivity of 
$BF\left( \mueee \right) < 5\times 10^{-15}$ at 90\% CL in a first phase, 
\mue[3] \phasei, using the existing $\rm{\pi E5}$ beamline at \psi. A 
further improvement is possible in a second phase, \mue[3] \phaseii, with 
upgrades to the muon beam (HiMB project) and detector to reach a sensitivity that is 
four orders of magnitude better than the current experimental limit.

The experimental signature of a \mueee\ decay at rest is given by three 
charged particle tracks, corresponding to the $e^{+}e^{-}e^{+}$ decay 
products, coincident in time, originating from a common vertex, and with a  
total energy consistent with the mass of the muon. Since this is a 
three-body decay, the energy of the decay products ranges from $< 1$~MeV 
up to a maximum of about half the mass of the muon.

The background has two main components: an intrinsic physics background 
coming from radiative muon decays (RMD) when the photon converts to an 
$e^{+}e^{-}$ pair; and an accidental background from the random 
combination of electrons and positrons from separate decays. The RMD 
background can be kept sufficiently small if the resolution on the 
$e^{+}e^{+}e^{-}$ energy sum can be kept below about $1$~MeV, while the 
suppression of the accidental background additionally relies on excellent  
timing and vertex resolution.

%%%%%%%%%%%%%%%%%%%%%%%%%%%%%%%%%%%%%%%%%%%%%%%%%%%%%%%%%%%%%%%%%%%%%%%%%%%%%%%%
\begin{sffamily}\begin{bfseries}
  {\textbf{Status and Plans of the \mue[3] experiment}}
\end{bfseries}\end{sffamily}
%%%%%%%%%%%%%%%%%%%%%%%%%%%%%%%%%%%%%%%%%%%%%%%%%%%%%%%%%%%%%%%%%%%%%%%%%%%%%%%%

The \mue[3] \phasei experiment~\cite{MU3E} is depicted in Fig.~2 of the Addendum. The 
main features include precision particle tracking with ultra-light, 
monolithic, silicon pixel tracking layers based on the HV-MAPS 
technology~\cite{ref:HVMAPS} cooled in an innovative manner using helium 
gas, plus a system of scintillating fibers and tiles to provide a time
resolution below the sub-nanosecond level, all located inside a 
superconducting solenoid providing a constant 1\,T field along the beam 
axis. A farm of GPUs will use a highly parallel 
track-fitting algorithm to perform full online tracking of all data as it 
streams continuously from the detector front-ends.  The detector is 
capable of handling very high rates and the \mue[3] \phasei  
sensitivity will be limited by the muon rate that \psi can deliver. A new 
Compact Muon Beamline has been installed for \mue[3] and will deliver a 
continuous high-purity beam of $28~\rm{MeV}/c$ muons at a rate of about
$10^8$~stop-$\mu/s$. The detector design is advanced and prototypes of 
all the main components have been built and successfully tested. 
Construction of the \mue[3] \phasei experiment will begin in 2019 and a 
first commissioning run is expected in 2020. After three years of 
operation the projected \phasei sensitivity is 
$BF\left( \mueee \right) < 5\times 10^{-15}$ at 90\% CL. However, the 
\mue[3] experimental concept allows for further significant improvements 
in sensitivity.
 
%%%%%%%%%%%%%%%%%%%%%%%%%%%%%%%%%%%%%%%%%%%%%%%%%%%%%%%%%%%%%%%%%%%%%%%%%%%%%%%%
\begin{sffamily}\begin{bfseries}
  {\textbf{Search for \mueee\ at future facilities}}
\end{bfseries}\end{sffamily}
%%%%%%%%%%%%%%%%%%%%%%%%%%%%%%%%%%%%%%%%%%%%%%%%%%%%%%%%%%%%%%%%%%%%%%%%%%%%%%%%

The sensitivity of \mue[3] \phasei is largely limited by the muon rate. 
The sensitivity to \mueee could be improved by more than an order of 
magnitude with a modest extension of the detector and access to a higher 
intensity muon beam. Roughly a factor two improvement in experimental 
acceptance is expected by extending the instrumentation in the forward 
and backward directions using the same pixel and scintillator 
technologies planned for \phasei.  At \psi, concepts for a new High 
Intensity Muon Beamline, \himb, have been 
investigated. Recent studies indicate that by refurbishing target M, and 
by installing a new capture solenoid the muon rate could be increase up to 
$1.3\times 10^{10}$~stop-$\mu^{+}/s$~\cite{Papa2018}, which is more than 
sufficient for \phaseii. Future facilities at \flab and \jparc could in 
principle provide similar muon intensities. If approved, the earliest \himb could be 
installed is 2024. After three years of data 
taking at the increased muon intensity the projected \phaseii sensitivity 
is
$BF\left( \mueee \right) < 1\times 10^{-16}$ at 90\% CL.

Detector R\&D should also continue to further improve the time resolution and
to reduce the amount of detector material, both being important 
requirements for suppressing accidental backgrounds at higher rates. 
Silicon pixel detectors with picosecond timing represent a
very promising technology for future \mue[3] upgrades.

\medskip
\begin{sffamily}\begin{bfseries}
  {\large The \muec\ process}
\end{bfseries}\end{sffamily}
%%%%%%%%%%%%%%%%%%%%%%%%%%%%%%%%%%%%%%%%%%%%%%%%%%%%%%%%%%%%%%%%%%%%%%%%%%%%%%%%

The \muec process is sensitive to new physics at mass scales up to
$10^{4}$~$\rm{TeV}/c^{2}$ and probes cLFV couplings that arise from
interactions where new physics appears in loop diagrams as well as from
$\mu e q q$ contact interactions. The most stringent limit on this process
was established by the \sindrumii experiment using data collected in
2000,
$R_{\mu e}\left( \mathrm{Au} \right) < 7\times 10^{-13}$
at $90\%$~CL~\cite{Mu2eLimit},
where, by convention,  $R_{\mu e}$ is the rate of the \muec\ conversion
process normalized to the normal muon nuclear capture process,
$\mu^{-} N \left( \mathrm{A, Z} \right) \rightarrow
\nu N^{*} \left( \mathrm{A, Z-1} \right)$.
Significant improvements in sensitivity offer genuine discovery 
possibilities across many new physics models, including several to which it is uniquely sensitive. 
%especially those for which \muegamma\ is suppressed. 
The \cometi experiment is currently under construction and aims to improve
sensitivity by a factor of 100 starting in the next few years. The \mue[2]
experiment is under construction and aims to improve the sensitivity
by four orders of magnitude by the mid 2020s, while \cometii is a proposed
upgrade to \cometi that would achieve a similar sensitivity or better. 
Further improvements are possible for both experiments with upgrades to the  
beamline and detectors.

The direct conversion process, \muec, is dominated by coherent
interactions with the nucleus to provide a two-body final state yielding a
clean experimental signature: an outgoing electron with an energy near the
muon mass (the recoil nucleus is not directly observed). The time
distribution of signal electrons should be consistent with the
characteristic lifetime of the muonic atoms formed as the initial $\mu^-$
beam comes to rest in a stopping target.

Significant sources of background events can arise from muons that
decay-in-orbit (DIO) --- that is, decay while captured in the atomic orbit
around a nucleus in the stopping target, from pions that survive to the
stopping target, and from cosmic ray muons that decay-in-flight or
interact in material to produce an electron with an energy near the muon
mass. If the energy of the initial proton beam is above the anti-proton
production threshold, then annihilations of the anti-protons can
contribute an additional source of background. The steeply falling DIO
background can be mitigated with excellent momentum resolution; the pion
background can be suppressed by using a pulsed
proton beam and employing a delayed live gate; the cosmic ray background
can be removed using a high-efficiency cosmic veto system; and anti-proton
backgrounds can be kept small by using absorbers to range out the
anti-protons far away from the stopping target.

Both the \comet and the \mue[2] experiments are based on a very clever
idea first proposed by Lobashov and Djilkabaev in 1989~\cite{MELC}. A
system of three solenoids --- a pion capture solenoid, a muon transport
solenoid, and a detector solenoid --- with graded magnetic fields which provide a 
significantly improved muon beam intensity to enable dramatic improvements
in sensitivity relative to \sindrumii. The feasibility of this method has
been demonstrated at low intensity at the MuSIC facility at the Research Center for Nuclear Physics in Osaka~\cite{MUSIC}.

%%%%%%%%%%%%%%%%%%%%%%%%%%%%%%%%%%%%%%%%%%%%%%%%%%%%%%%%%%%%%%%%%%%%%%%%%%%%%%%%
\begin{sffamily}\begin{bfseries}
  {\textbf{Status and Plans of the \comet experiment}}
\end{bfseries}\end{sffamily}
%%%%%%%%%%%%%%%%%%%%%%%%%%%%%%%%%%%%%%%%%%%%%%%%%%%%%%%%%%%%%%%%%%%%%%%%%%%%%%%%

The \cometi experiment~\cite{COMET-I-TDR} is depicted in Fig.~4 of the Addendum. The
apparatus begins with a pion-production target made of graphite located inside the pion 
capture solenoid, which provides a graded magnetic field to collect 
low-momentum pions by reflecting them backwards with respect to the 
incoming proton beam. The muon transport solenoid is a curved 90-degree 
magnet that, together with a set of dipole coils, serves as a tranport and 
charge- and momentum-selection channel for $\pi^{-}\rightarrow\mu^{-}\nu$ decays. The 
muon transport solenoid delivers a high-intensity $\mu^-$ beam to the 
detector solenoid, which houses an aluminum stopping target and active 
detector elements, including a cylindrical drift chamber (CDC) for the \muec\ search and low-mass straw chambers and a 
fast LYSO crystal calorimeter for beam measurements. An active cosmic-ray veto system shadows the 
detector and stopping target regions outside the solenoid volume. 
Additional instrumentation monitors the proton and muon beams. The  
construction of the entire apparatus is at an advanced stage, with two of 
the magnets and the CDC complete, including significant European 
contributions to the cosmic-veto, muon stopping target and beam monitoring systems, the 
trigger and DAQ, and computing and software. Beam commissioning is 
expected to begin in 2020. The \cometi experiment will utilize about
3~kW of 8~GeV protons from the \jparc Main Ring, delivered in pulses
spaced by $1.17$~$\mu s$, to first make
important measurements of the muon yield and determine rates for various
background processes before concentrating on a search for cLFV. After 150
days of operation the projected \cometi sensitivity is
$R_{\mu e} < 7\times 10^{-15}$ at $90\%$~CL~\cite{COMET-I-TDR}. This
sensitivity can be significantly improved using a higher intensity proton
beam and extending the \cometi apparatus.

%%%%%%%%%%%%%%%%%%%%%%%%%%%%%%%%%%%%%%%%%%%%%%%%%%%%%%%%%%%%%%%%%%%%%%%%%%%%%%%%
\begin{sffamily}\begin{bfseries}
  {\textbf{Status and Plans of the \mue[2] experiment}}
\end{bfseries}\end{sffamily}
%%%%%%%%%%%%%%%%%%%%%%%%%%%%%%%%%%%%%%%%%%%%%%%%%%%%%%%%%%%%%%%%%%%%%%%%%%%%%%%%

The \mue[2] experiment~\cite{MU2E} is depicted in Fig.~6 of the Addendum. The graded high-field 
pion production solenoid collects and focuses low-momentum pions towards 
the muon transport solenoid, which is an ``S''-shaped magnet with a total 
path length of about 13 meters. The muon transport solenoid includes a set 
of collimators for momentum and charge-selection to provide $\sim 
10^{10}$~stop-$\mu^{-}/s$ using 8~kW of 8~GeV protons from the \flab 
Booster delivered in pulses spaced $1.7$~$\mu s$ apart. The detector 
solenoid  provides a graded magnetic field in the upstream
region, which houses the stopping target, and a near constant magnetic 
field in the downstream region, which houses the active detector 
elements. A low-mass tracking system consisting of approximately 21k 
thin aluminized-mylar straws~\cite{Mu2e-tracker} and a calorimeter 
consisting of two annular disks of pure CsI 
crystals~\cite{Mu2e-calorimeter} precisely measure 
the timing, energy, and momenta of particles originating from the stopping 
target. The apparatus is shadowed on the outside by a large, 
scintillator-based cosmic-veto system. Ancillary systems are used to 
monitor the quality and intensity of the proton and muon beams. Construction of the 
solenoids and all the detector sub-systems has begun, with significant 
European contributions to the muon transport solenoid, the calorimeter, 
and the muon beam monitoring system. Commissioning is expected to begin in 
2022. After 690 days of operation the projected sensitivity is
$R_{\mu e} < 8\times 10^{-17}$ at $90\%$~CL~\cite{MU2E}. This 
sensitivity can be improved by at least a factor of 10 using a higher 
intensity proton beam and upgrading the \mue[2] apparatus.

%%%%%%%%%%%%%%%%%%%%%%%%%%%%%%%%%%%%%%%%%%%%%%%%%%%%%%%%%%%%%%%%%%%%%%%%%%%%%%%%
\begin{sffamily}\begin{bfseries}
  {\textbf{Search for \muec\ at future facilities}}
\end{bfseries}\end{sffamily}
%%%%%%%%%%%%%%%%%%%%%%%%%%%%%%%%%%%%%%%%%%%%%%%%%%%%%%%%%%%%%%%%%%%%%%%%%%%%%%%%

The \cometi experiment can be extended, as depicted in Fig. 5 of the
Addendum, and utilize 56~kW of 8~GeV protons from the \jparc Main
Ring, delivered in pulses spaced by 1.17~$\mu s$, to reach a
sensitivity of $R_{\mu e} < 2.6 \times 10^{-17}$ at $90\%$~CL with
about 230~days of operation. Further improvements by one order of
magnitude from refinements to the experimental design and operation
are being considered, within the beam power and the beam time as
originally assumed \cite{COMET-ESPP}.  These improvements include
dipole steering fields in the curved muon transport and electron
spectrometer sections to allow a more fine-tuned momentum selection
which is important to optimize the acceptance and background
rejection. The detailed measurements from \phasei will provide
important input to the final \phaseii design and construction. Data
taking could begin in the mid-2020s.

The \mue[2] experiment can be upgraded, \mueii, to take advantage of the 
increased 
proton beam intensity available from the \pipii project, currently in the 
design phase at \flab.  The \pipii linac is expected to become 
operational in the latter half of the 2020s and will provide 1.6~MW of 
0.8~GeV protons with a programmable time structure. An Expression of 
Interest for \mueii \cite{Mu2e-II-EOI} was recently submitted to the 
Fermilab Physics Advisory Committee, which concluded that the science 
case was compelling and recommended that funding for high-priority R\&D 
be identified. The Expression of Interest included signatures from 130 
scientists from 36 institutions in six countries, including Italy, 
Germany, and the UK. Using 100~kW of protons from \pipii, the \mueii 
projected sensitivity is a factor ten or more better than the \mue[2]  
sensitivity. Data taking could begin in the late 2020s.

The \comet collaboration is also heavily involved in R\&D towards the 
\prism\ project, which combines \cometii with an FFAG muon storage ring to potentially provide muon beam intensities 
of $>10^{12}$~stop-$\mu /s$ with a narrow momentum bite allowing the use of 
very thin stopping targets, and significantly reduced pion contamination 
owing to the increased transport path length. In conjunction with an 
upgrade to the \jparc proton source to achieve 1.3~MW and to the detector  
systems to accomodate the higher rates, \prism\ offers the potential to 
achieve  sensitivies to \muec\ of the order of $10^{-19}$. The  
monochromatic, pion-suppressed, high-intensity muon beam provided by \prism\ 
will allow the use of stopping targets comprised of heavy elements, such 
as gold or lead, that can be important in understanding the underlying new 
physics operators in the event of a discovery~\cite{COMET-ESPP}.

\bigskip
\begin{sffamily} \begin{bfseries}
{\Large Summary}
\end{bfseries} \end{sffamily}

The \meg, \mue[3], \mue[2], and \comet experiments use intense muon 
beams to provide the broadest, deepest, most sensitive probes of 
charged-lepton flavour violating interactions and to explore the BSM parameter 
space with sensitivity to new physics mass scales of 
$10^3 - 10^4$~$\mathrm{TeV}/c^2$,
well beyond what can be directly probed at colliders. Over the next five years,
currently planned experiments in Europe, the US, and Asia will begin
taking data and will extend the sensitivity to $\mu\rightarrow e$ 
charged-lepton flavour violating transitions by orders of magnitude. Further 
improvements are possible and new or upgraded experiments are
being considered that would utilize upgraded accelerator facilities at \psi,
\flab, and \jparc. The schedule of planned and proposed experiments 
is summarized in the figure below. Strong European participation in the 
design, construction, data taking, and analysis will be important for the 
success of these future endeavors and represents a prudent investment complementary to searches at colliders.

We urge the committee to strongly support the continued participation of
European institutions in experiments searching for charged-lepton flavour 
violating $\mu\rightarrow e$
transitions using high-intensity beams at facilities in Europe, the US, and 
Asia, including possible upgraded experiments at next-generation facilities 
available the latter half of the next decade at \psi, \flab, and \jparc.

\begin{figure}[h!]
\begin{center}
  \includegraphics[width=0.99\textwidth]{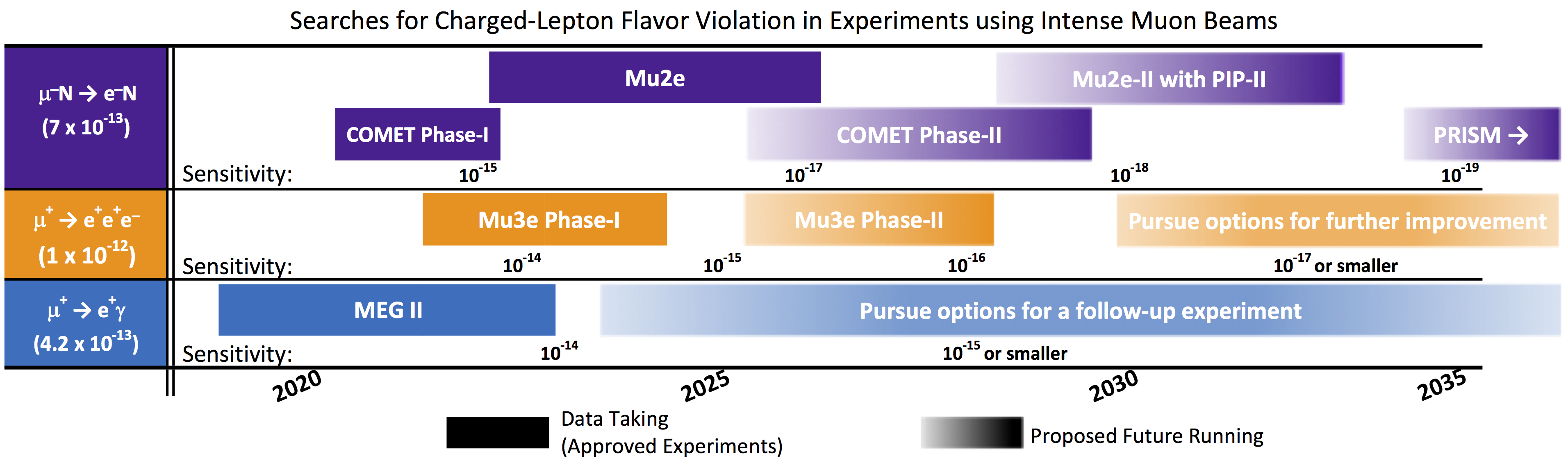}
\end{center}
\caption{\it{\small{Planned data taking schedules for current experiments that search for charged-lepton flavor violating 
$\mu\rightarrow e$  transitions. Also shown are possible schedules 
for future proposed upgrades to these experiments. The current 
best limits for each process are shown on the left in parentheses, while 
expected future sensitivities are indicated by order of magnitude 
along the bottom of each row.}}} 
\label{fig:Schedule} 
\end{figure}

\newpage
\pagenumbering{arabic}
\setcounter{page}{1}
\setcounter{figure}{0}

\mbox{}
\vskip 3 cm

\begin{center} \begin{sffamily} \begin{bfseries}
  {{\Huge Addendum:} \\
  \vskip 0.25 cm

   {\Large Charged Lepton Flavour Violation using \\  
     Intense Muon Beams at Future Facilities}}
\end{bfseries} \end{sffamily} \end{center}
\vskip 1 cm

\begin{center}
{\large A. Baldini,
  D. Glenzinski,
  F. Kapusta,
  Y. Kuno,
  M. Lancaster, \\
  J. Miller,
  S. Miscetti,
  T. Mori,
  A. Papa,
  A. Sch\"oning,
  Y. Uchida}
\end{center}
\vskip 1 cm

\begin{center}
  {\large
A submission to the 2020 update of the European Strategy for Particle 
Physics on behalf of the COMET, MEG, Mu2e and Mu3e collaborations. }
\end{center}
\vskip 2 cm

\begin{center} \begin{sffamily} \begin{bfseries}
  {\Large Abstract}
\end{bfseries} \end{sffamily} \end{center}

In this Addendum additional information is provided about the MEG, Mu3e, 
Mu2e, and COMET experiments and their associated collaborations. The 
contributions from Europe are emphasized.
\vskip 8cm
Contact: Andr\'e Sch\"oning [schoning@physi.uni-heidelberg.de] 

% --- cover page for the Addendum

\newpage 
\mbox{}
\vskip 1 cm

\begin{sffamily}\begin{bfseries}
  {\Large Addendum for the MEG Experiment}
\end{bfseries}\end{sffamily}
%%%%%%%%%%%%%%%%%%%%%%%%%%%%%%%%%%%%%%%%%%%%%%%%%%%%%%%%%%%%%%%%%%%%%%%%%%%%%%%%

%
\begin{figure}[htbp]
\begin{center}
  \includegraphics[width=0.7 \textwidth]{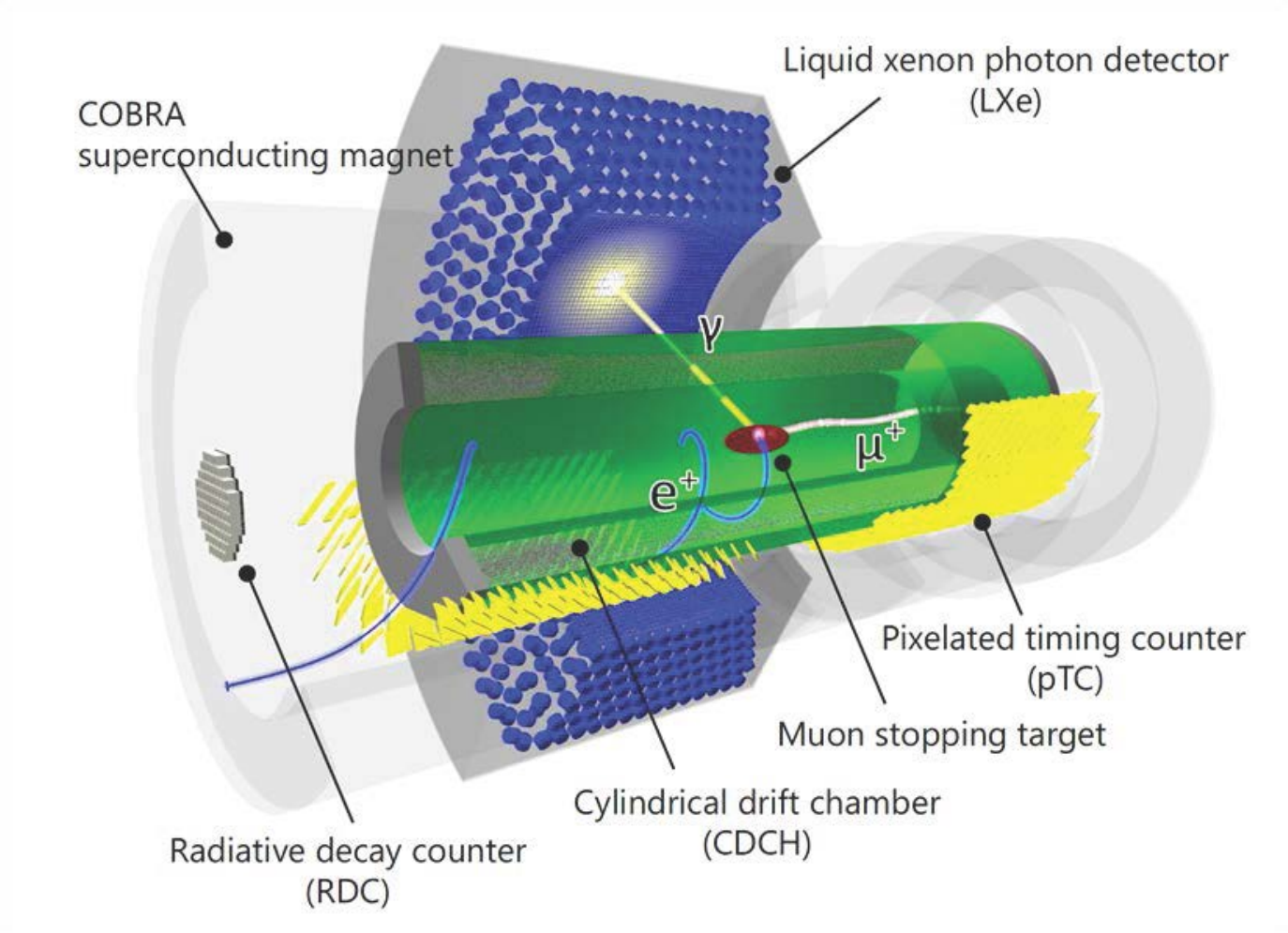}
\end{center}
\caption{\it{\small{Schematic of the \megii experiment.}}}
\label{fig:MEG}
\end{figure}
%

%%%%%%%%%%%%%%%%%%%%%%%%%%%%%%%%%%%%%%%%%%%%%%%%%%%%%%%%%%%%%%%%%%%%%%%%%%%%%%%%

\begin{sffamily}\begin{bfseries}
  {\textbf{Experiment Website and Contact Information}}
\end{bfseries}\end{sffamily}

Website: http://meg.web.psi.ch \\
Co-spokespersons: 
A. Baldini (University of Pisa) and T. Mori (University of Tokyo) \\
\hspace*{3.3cm}(alessandro.baldini@pi.infn.it, mori@icepp.s.u-tokyo.ac.jp)

\begin{sffamily}\begin{bfseries}
  {\textbf{Interested Community}}
\end{bfseries}\end{sffamily}

The \megii Collaboration consists of about 75 participants from 
Japanese, Italian, Swiss,  Russian and US  institutions. Scientists and 
students from Europe account for 50\% of the collaboration. The experiment 
is hosted at the \psi laboratory in Switzerland.

\begin{sffamily}\begin{bfseries}
  {\textbf{Timeline}}
\end{bfseries}\end{sffamily}

The \megii experiment has recently completed construction and first 
commissioning data was collected in 2018. A three year physics run 
is expected to begin in 2019.

\begin{sffamily}\begin{bfseries}
  {\textbf{European Contributions}}
\end{bfseries}\end{sffamily}

The European contributions to \megii spanned all the major 
sub-sytems of the experiment including:
\begin{itemize}
\item{Data acquisition software}
\item{Construction of trigger and read-out electronics}
\item{Procurement of silicon photomultipliers for the positron timing counter}
\item{Mechanical structure of the positron timing counter}
\item{Construction of the new cylindrical drift chamber}
\item{Construction of the  liquid xenon detector cryostat}
\item{Calibration devices for the liquid xenon detector}
\end{itemize}
The European groups also play a significant role in the leadership, 
commissioning, operations, analysis, and publication activities of the 
experiment.

\begin{sffamily}\begin{bfseries}
  {\textbf{Computing Requirements}}
\end{bfseries}\end{sffamily}

The computing system of  \megii\ consists in about 320 CPU cores 
and 1300/2000~TB of disk/tape space.
Computing expenses are equally subdivided among Japanese, Italian and 
Swiss institutions.

%%%%%%%%%%%%%%%%%%%%%%%%%%%%%%%%%%%%%%%%%%%%%%%%%%%%%%%%%%%%%%%%%%%%%%%%%%%%%%%%

\newpage 
\mbox{}
\vskip 1 cm

\begin{sffamily}\begin{bfseries}
  {\Large Addendum for the Mu3e Experiment}
\end{bfseries}\end{sffamily}
%%%%%%%%%%%%%%%%%%%%%%%%%%%%%%%%%%%%%%%%%%%%%%%%%%%%%%%%%%%%%%%%%%%%%%%%%%%%%%%%

%
\begin{figure}[htbp]
\begin{center}
  \includegraphics[width=0.8\textwidth]{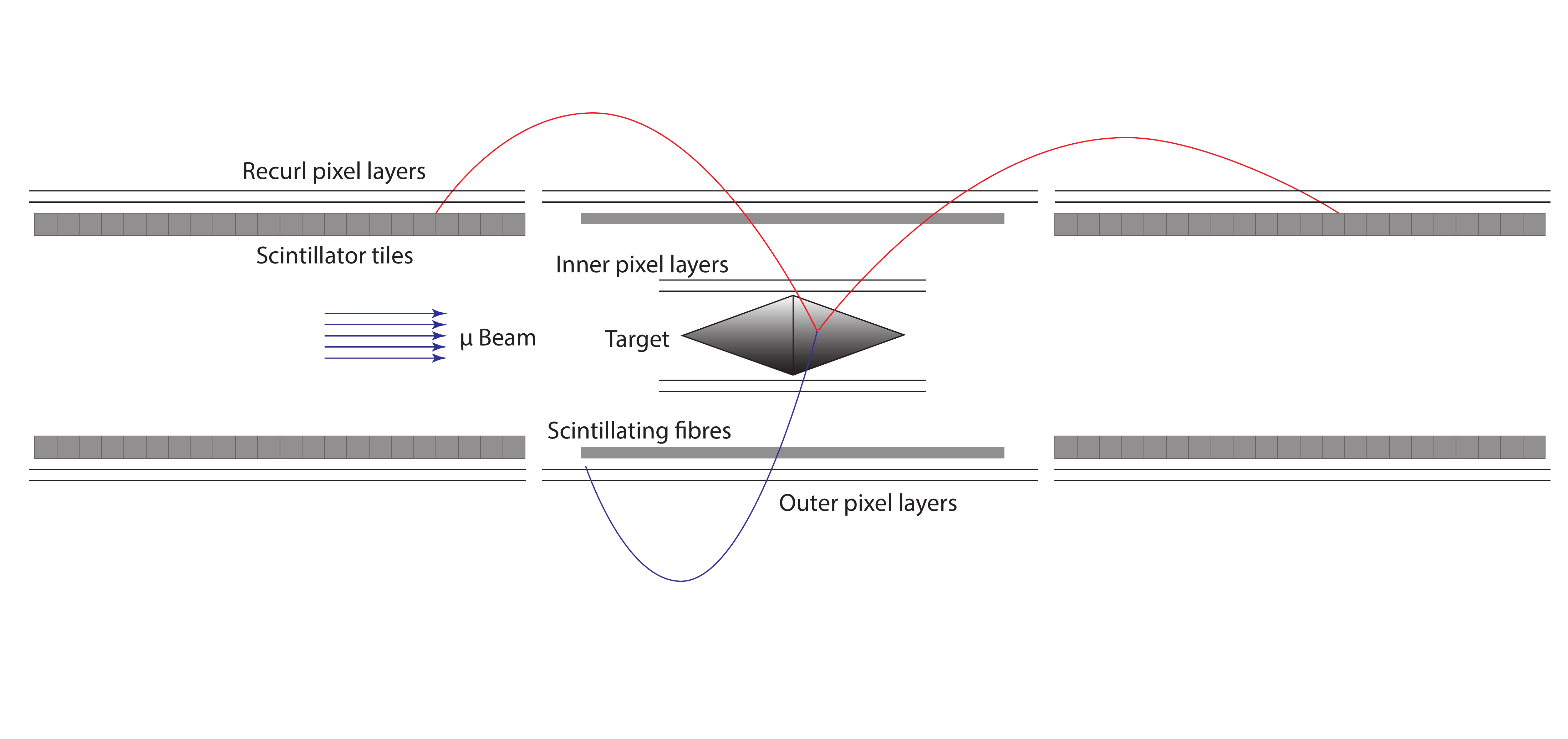}
\end{center}
\vskip -1.5cm
\caption{\it{\small{Schematic of the \mue[3] experiment.}}}
\label{fig:Mu3e}
\end{figure}
%

%%%%%%%%%%%%%%%%%%%%%%%%%%%%%%%%%%%%%%%%%%%%%%%%%%%%%%%%%%%%%%%%%%%%%%%%%%%%%%%%

\begin{sffamily}\begin{bfseries}
  {\textbf{Experiment Website and Contact Information}}
\end{bfseries}\end{sffamily}

Website: https://www.psi.ch/mu3e/mu3e \\
Co-spokespersons: 
S. Ritt (PSI Laboratory) and A. Sch\"{o}ning (Heidelberg University)  \\
\hspace*{3.3cm}(stefan.ritt@psi.ch, schoning@physi.uni-heidelberg.de)

\begin{sffamily}\begin{bfseries}
  {\textbf{Interested Community}}
\end{bfseries}\end{sffamily}

The \mue[3] Collaboration consists of about 70 participants, from eleven  
European institutions in Germany, Switzerland and United Kingdom. Scientists 
and Europe account for 100\% of the collaboration. The experiment is hosted 
at the \psi laboratory in Switzerland.

\begin{sffamily}\begin{bfseries}
  {\textbf{Timeline}}
\end{bfseries}\end{sffamily}

The experiment will be performed in two phases. The R\&D programme 
is nearly complete and construction has begun for various components. 
Commissioning with beam for \mue[3] \phasei is expected to start in 2020.
Three years of physics data taking is required to reach the design 
sensitivity.

\begin{figure}[htbp]
\begin{center}
\includegraphics[width=0.89\linewidth]{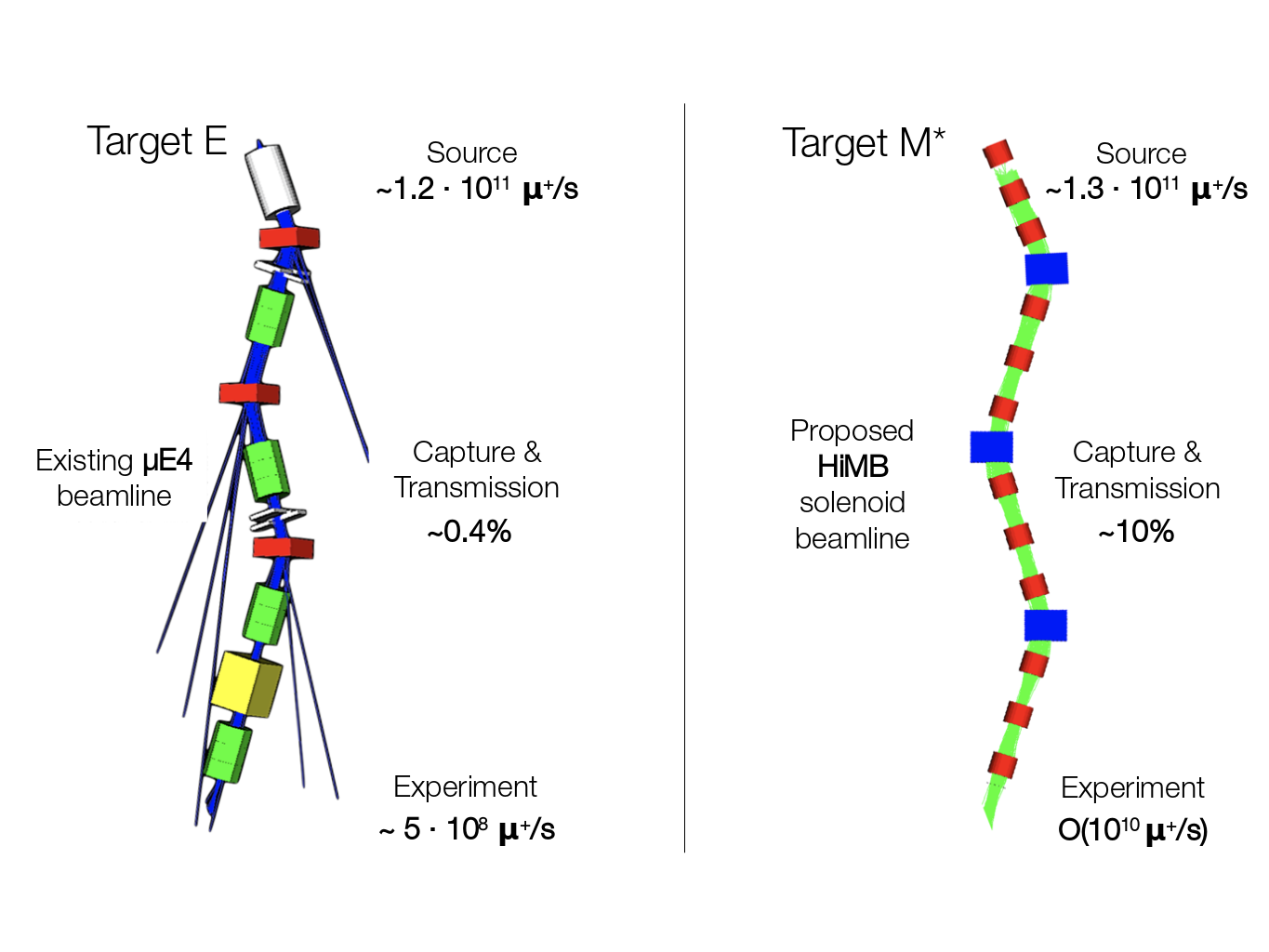}
\caption{The new proposed solenoidal beamline for \himb\ (right) compared 
to the current hybrid $\mu$E4 beamline (left).}
\label{fig:HiMB}
\end{center}
\end{figure}

The \mue[3] \phaseii experiment requires an upgraded detector with an 
extended geometrical acceptance and the construction of a new high 
intensity muon beamline, \himb, at \psi. 
The proposal requires refurbishing target M of the proton beamline and
installation of a new capture solenoid followed by a solenoidal beamline,
using a design similar to existing $\mu${E4} beamline, see Fig.~\ref{fig:HiMB}.
Ongoing studies are investigating whether, with modest 
modifications, the 
\phaseii experiment may also allow searches for \muegamma\ decays or  
muonium-anitmuonium oscillations.
Design studies for \himb are 
underway and installation, if approved, could start at the earliest in 
2024 after the completion of the \phasei programme.

\begin{sffamily}\begin{bfseries}   
  {\textbf{European Contributions}}
\end{bfseries}\end{sffamily}

The entire \mue[3] \phasei experiment is designed and built by European 
institutions.  The main components of the experiment are:
\begin{itemize}
  \item{Superconducting solenoid with a homogeneous magnetic field of 
     $B=1~\textrm{Tesla}$ (up to B=2.6~\textrm{Tesla}).}
  \item{Ultra-light pixel tracker based on high voltage monolithic active 
     pixel sensors (HV-MAPS).}
  \item{Two scintillating detector systems for sub-nanosecond timing, based on
     scintillating fibers and tiles.}
  \item{Trigger-less data acquisition system with continuous readout.}  
  \item{Filter farm based on graphical processing units.}
\end{itemize}
The European groups also play a significant role in all aspects of the 
experiment including leadership, operations, analysis and publication 
activities.

Most groups of the \mue[3] collaboration \phasei have expressed interest to
contribute to \phaseii. Also new groups are invited to contribute to the
planned \mue[3] \phaseii upgrade, and to investigate further extensions of
the \mue[3] physics programme, for example a search for \muegamma with a  photon
conversion layer or  
muonium-antimuonium oscillations with an upgraded \mue[3] detector.

\begin{sffamily}\begin{bfseries}
  {\textbf{Computing Requirements}}            
\end{bfseries}\end{sffamily}

The computing system and needs will be similar to those of the MEG
experiment. Expenses for computing will be shared among the contributing
institutes. Additional GRID computing resources will be required to fully
exploit the physics potential of the experiment.

%%%%%%%%%%%%%%%%%%%%%%%%%%%%%%%%%%%%%%%%%%%%%%%%%%%%%%%%%%%%%%%%%%%%%%%%%%%%%%%%

\newpage 
\mbox{}
\vskip 1 cm

\begin{sffamily}\begin{bfseries}
  {\Large Addendum for the COMET Experiment}
\end{bfseries}\end{sffamily}
%%%%%%%%%%%%%%%%%%%%%%%%%%%%%%%%%%%%%%%%%%%%%%%%%%%%%%%%%%%%%%%%%%%%%%%%%%%%%%%%

% --- COMET Phase-I figure
\begin{figure}[htbp]
  \begin{center}
    \includegraphics[width=0.6\textwidth]{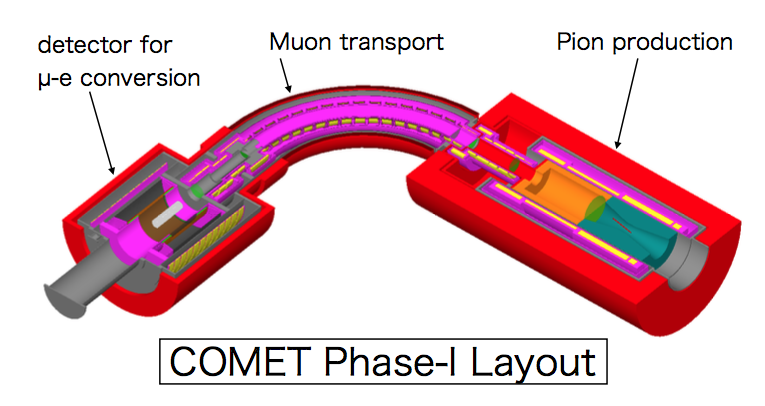}
  \end{center}\label{fig:comet}
  \caption{\it{\small{A schematic of the \cometi experiment. A cosmic-ray 
    veto system and monitors for the proton beam and muon beam are not 
    shown.}}}
  \label{fig:cometi}  
\end{figure}
%

%%%%%%%%%%%%%%%%%%%%%%%%%%%%%%%%%%%%%%%%%%%%%%%%%%%%%%%%%%%%%%%%%%%%%%%%%%%%%%%%

\begin{sffamily}\begin{bfseries}
  {\textbf{Experiment Website and Contact Information}}
\end{bfseries}\end{sffamily}

Website: http://comet.kek.jp/Introduction.html \\
Spokesperson: Y. Kuno (Osaka University) \\
\hspace*{2.7cm}(kuno@phys.sci.osaka-u.ac.jp)

\begin{sffamily}\begin{bfseries}
  {\textbf{Interested Community}}
\end{bfseries}\end{sffamily}

The \comet Collaboration consists of about 200 participants from 35 
institutions from Australia, Belarus, China, Czech Republic, France, 
Georgia, Germany, India, Japan, Kazakhstan, South Korea, 
Malaysia, Russia, United Kingdom, and Vietnam. Scientists and students from 
Europe account for about 30\% of the collaboration. The experiment is 
hosted at the \jparc laboratory in Japan.

\begin{sffamily}\begin{bfseries}
  {\textbf{Timeline}}
\end{bfseries}\end{sffamily}

The experiment will be performed in two phases. Construction of \cometi is 
at and advanced stage. The \jparc proton beam will arrive at the \comet experimental 
area in early 2020, when \phasei beam studies and integration will commence.
The \phasei physics data-taking and analysis will follow.

The \comet\ \phaseii experiment requires the construction of an extended 
solenoid system as depicted in Fig.~5. that, if approved, could be ready 
in the mid--2020s.
The completed \cometii configuration can be adapted to search for and 
measure several charged-lepton flavour- and number-violating (cLNV) 
processes other than the main \muec channel, and a broad programme of 
study is expected to continue well beyond 2025 and into the 2030s, with a 
specific path that is dependent on the observations that have been made by 
that time.
Some of these additional measurements will require the beamline to run in 
dedicated positive-muon mode, which will produce an extremely high-quality 
beam in the \phaseii configuration.

In the longer term (2030 and beyond), the \comet collaboration is 
also closely engaged with the next-generation PRISM experiment through the 
PRISM Task Force, which makes use of an FFAG 
muon storage ring to pursue detailed measurements of cLFV and LNV 
processes. This is a relatively 
long-term project which would be expected in the latter stages of the 
period relevant to the present strategy exercise.

\begin{sffamily}\begin{bfseries}
  {\textbf{European Contributions}}            
\end{bfseries}\end{sffamily}

The European contributions to \comet include: 
\begin{itemize}
\item{Cosmic Ray Veto detector (Belarus, France, Georgia, Russia)}
\item{Electromagnetic calorimeter (Belarus, Russia)}
\item{Muon target monitor (Germany)}
\item{Data-acquisition and detector triggering systems (UK, Czech Republic)}
\item{Straw-tube tracking detector (Georgia, Russia)}
\item{Muon stopping targetry (Germany)}
\end{itemize}
The European groups also play a significant role in the leadership, 
analysis, and publication activities and are expected to play a significant 
role in the commissioning and operation of the experiment beginning in 
2020. The international PRISM task force also benefits from significant 
European contributions, including leadership.

\begin{sffamily}\begin{bfseries}
  {\textbf{Computing Requirements}}            
\end{bfseries}\end{sffamily}

Controlling and monitoring the beam composition and the various 
backgrounds for this rare-decay experiment requires very large simulated 
data samples. Single- and multi-bunch simulations have involved 
significant contributions in terms of CPU (France, UK and Germany), 
storage and data sharing (France). Software developments related to the 
analysis, track finding and track fitting optimization lead also to 
intensive software tests and improvements (UK, France, Germany). In 
particular, much effort has been focused on introducing simulation 
strategies that allow for high-statistics background and signal estimates 
without requiring a proportional increase in computational resources. 
Combining such strategies with increasing international resource 
contributions will allow the computational challenges of \comet to be met.

% --- COMET Phase-II Figure:

\begin{figure}[htbp]
  \begin{center}
    \includegraphics[width=0.6\textwidth]{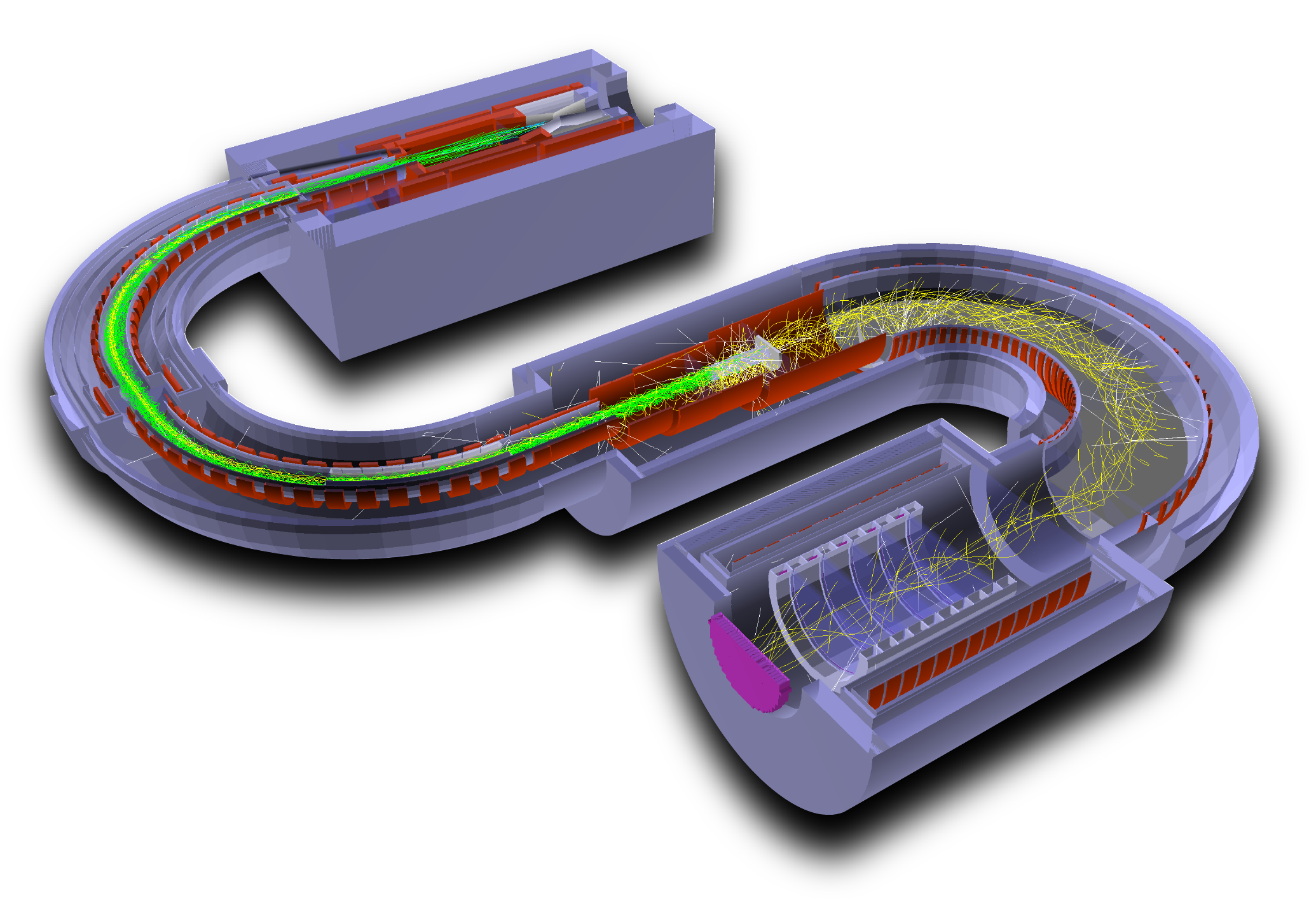}
  \end{center}
  \caption{\it{\small{Schematic of the \cometii experiment.}}}
  \label{fig:cometii}
\end{figure}

\newpage 
\mbox{}
\vskip 1 cm

\begin{sffamily}\begin{bfseries}
  {\Large Addendum for the Mu2e and Mu2e-II Experiment}
\end{bfseries}\end{sffamily}
%%%%%%%%%%%%%%%%%%%%%%%%%%%%%%%%%%%%%%%%%%%%%%%%%%%%%%%%%%%%%%%%%%%%%%%%%%%%%%%%

%
\begin{figure}[htbp]
\begin{center}
  \includegraphics[width=0.8\textwidth]{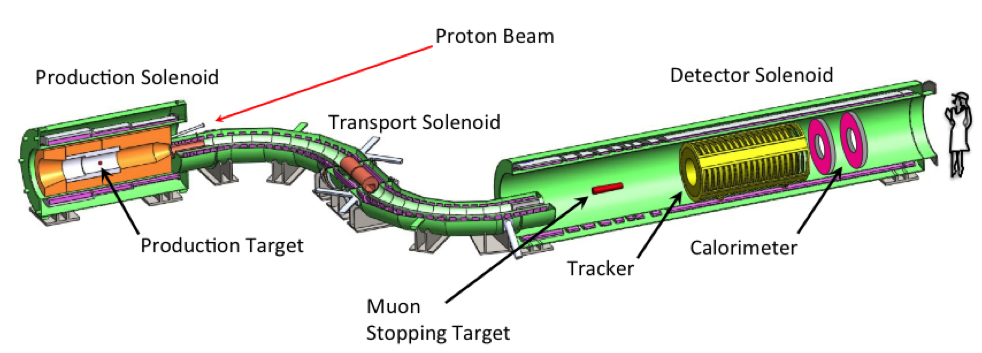}
\end{center}
\caption{\it{\small{Schematic of the \mue[2] experiment. A cosmic-ray veto 
system, and monitors for the proton beam and muon beam are not shown.}}}
\label{fig:Mu2e}
\end{figure}
%

%%%%%%%%%%%%%%%%%%%%%%%%%%%%%%%%%%%%%%%%%%%%%%%%%%%%%%%%%%%%%%%%%%%%%%%%%%%%%%%%

\begin{sffamily}\begin{bfseries}
  {\textbf{Experiment Website and Contact Information}}
\end{bfseries}\end{sffamily}

Website: https://mu2e.fnal.gov \\
Co-spokespersons: 
D. Glenzinski (Fermilab) and J. Miller (Boston University) \\
\hspace*{3.3cm}(mu2e-spokespersons@fnal.gov)

\begin{sffamily}\begin{bfseries}
  {\textbf{Interested Community}}
\end{bfseries}\end{sffamily}

The \mue[2] Collaboration consists of 242 members from 40 institutions in 
China, Germany, Italy, Russia, the United Kingdom, and the United States. 
Scientists and students from European institutions account for 26\% of the 
collaboration. The experiment is hosted by \flab in the US.

\begin{sffamily}\begin{bfseries}
  {\textbf{Timeline}}
\end{bfseries}\end{sffamily}

The \mue[2] experiment is currently under construction. In 2021 
commissioning of the proton beamline, and cosmic-ray commissioning of the 
detector systems are scheduled to begin. Commissioning of the detector 
systems with beam is expected in 2022 and a four-year physics run is 
planned starting in 2023.

In parallel to \mue[2] construction and commissioning, R\&D for \mueii
will begin in order to develop a conceptual design for the detectors and a 
new proton beamline to accommodate the new beam energy provided by the 
\pipii linac. There are challenging issues associated with the increased 
rate and radiation environment for the production solenoid, the production 
target, and the detector systems and their associated read-out electronics.
The timeline for \mueii will be driven by the completion of \mue[2] as well 
as the construction of the \pipii linac, which, if approved, is expected 
to become available in the mid-2020 timescale. To achieve another factor 
of ten or more improvement in sensitivity will require about three years 
of physics data taking with 100~kW of protons from \pipii. The flexibility 
of \pipii provides an opportunity to deliver customized muon beams for 
the exploration of other \mueii stopping target materials as well as 
for next-generation \mueee\ or \muegamma\ experiments.

\begin{sffamily}\begin{bfseries}
  {\textbf{European Contributions}}            
\end{bfseries}\end{sffamily}

The European contributions to \mue[2] spanned several important 
sub-systems of the experiment including:
\begin{itemize}

\item{Calorimeter: the design and construction is lead by Italy with 
additional contributions from Germany, Russia, and the US. Italy (INFN) 
is contributing $\mathcal{O}$(3M Euro) for core construction costs and 
provided support for $\mathcal{O}$(30) people.} 
\item{Muon Target Monitor: the final design and construction of the muon 
target monitor is driven by the UK in collaboration with the US. The UK 
(STFC) is contributing $\mathcal{O}$(1M Euro) for core construction costs 
and provided support for $\mathcal{O}$(15) people.} 
\item{Transport Solenoid: Italy made very significant contributions to the 
design, prototyping, and fabrication of the superconductor and coils of 
the transport solenoid.}
\item{Irradiation facilities: Germany provides support for 
$\mathcal{O}$(2) people plus in-kind access to the EPOS 
and G-ELBE irradiation facilities at HZDR for tests of the rate 
capabilities and radiation tolerance of various detector sub-systems.} 
\end{itemize}
European groups also play a significant role in the leaderhip, analysis, 
and publication activities and are expected to play a significant role in 
the commissioning and operation of the experiment beginning in 2021.

For \mueii, European groups have expressed interest in contributing to the 
development and design of an upgraded calorimeter (e.g. using BaF$_2$ 
crystals and optimized photosensors), of upgraded read-out electronics 
using next-generation FPGAs or custom ASICs, and of an upgraded tracker 
potentially using micro-RWell or MPGD technologies.

\begin{sffamily}\begin{bfseries}
  {\textbf{Computing Requirements}}            
\end{bfseries}\end{sffamily}

The computing resources required for \mue[2] data processing, 
reconstruction, and analysis are estimated to be about 1000 CPU cores and 
30/60~PB of disk/tape space. Significant additional CPU resources 
($\sim$\,30M CPU-hours per year) from the WLCG and high performance computing 
systems (e.g. NERSC) are utilized annually to produce high-statistics 
simulation samples.

% --- END OF DOCUMENT

\newpage
\begin{thebibliography}{}
\bibitem{BFmeg-petcov} S.T. Petcov, Sov. J. Nucl. Phys. 25 (1977) 340.
\bibitem{muoniumoscillations} L. Willmann, {\it{et al.}}, Phys.Rev.Lett. 
  82 (1999) 49.
\bibitem{CalibiSignorelli} L. Calibbi and G. Signorelli, 
  Riv. Nuovo. Cimento, 41 (2018) 71.
\bibitem{Cirigliano} V. Cirigliano, {\it{et al.}}, Phys. Rev. D80 (2009) 
  013002.
\bibitem{MEGLimit} A. Baldini, {\it{et al.}} ({\meg} Collaboration), 
  Eur. Phys. J. C76 (2016) 434.
\bibitem{Mu3eLimit} U. Bellgardt, {\it{et al.}} ({\sindrum} Collaboration),
  Nucl. Phys.  B299 (1988) 1.
\bibitem{Mu2eLimit} W. Bertl, {\it{et al.}} ({\sindrumii} Collaboration),  
  Eur. Phys. J. C47 (2006) 337.
\bibitem{BNL-PRD-2004} G.W. Bennett, {\it{et al.}} ({\tt{E821}} 
  Collaboration), Phys. Rev. Lett. 92 (2004) 161802.
\bibitem{gm2_CLFV}  G. F. Giudice, P. Paradisi and M. Passera, JHEP11 (2012) 113.
\bibitem{SEESAWS}  T. Hambye, Proc. Nucl. Phys. B248 (2014) 13.
\bibitem{INV-SS} A. Abada, {\it{et al.}}, JHEP 11 (2014) 048.
\bibitem{MuminusToEplus} J. Kaulard, {\it{et al.}} ({\sindrumii} 
  Collaboration), Phys. Lett. B422 (1998) 334.
\bibitem{MuminusToEplusProspects} B.~Yeo, Y.~Kuno, M.~Lee and K.~Zuber,
%%%{\it{Future experimental improvement for the search of 
%%%lepton-number-violating processes in the eμ sector}},
Phys.Rev. D96, no. 7 (2017) 075027. 
\bibitem{LQ} B. Gripaios, {\it{et al.}}, JHEP (2015) 6; 
  B. Dumont, {\it{et al.}}, arXiv:1603.05248 (2016); 
  M. Bauer and M. Neubert, Phys. Rev.Lett. 116 (2016) 141802; 
  S. Baek and K. Nishiwaki, Phys. Rev. D93 (2016) 015002.
\bibitem{Z-PRIME} A. Crivellin, {\it{et al.}}, Phys. Rev. D92 (2015) 050413.  
\bibitem{LQ-CLFV} J. Arnold, {\it{et al.}}, Phys. Rev. D88 (2013) 035009.
\bibitem{LFV-LHCB} R. Aaij, {\it{et al.}} ({\tt{LHCb}} Collaboration), 
  Phys.Rev.Lett. 111 (2013) 141801.
\bibitem{tauFV} G. Wilkinson {\it{et al.}}, {\tt https://indico.cern.ch/event/706741/contributions/3017537}  
\bibitem{PIPII} M.Ball, {\it{et al.}} (PIP-II Accelerator Facility), \\
 {\tt http://pip2-docdb.fnal.gov/cgi-bin/ShowDocument?docid=113 (2018).}
\bibitem{MEGII} A. Baldini, {\it{et al.}} ({\megii} Collaboration), 
  Eur. Phys. J. C78 (2018) 380.
\bibitem{Papa} A. Papa, {\it{et al.}}, Nucl. Phys. Proc. Suppl. 248 (2014) 121.
\bibitem{Cavoto} G. Cavoto, {\it{et al.}}, Eur.Phys.J. C78 (2018) 37.
\bibitem{Hitlin} C. Cheng, B. Echenard, D.G. Hitlin, arXiv:1309.7679
(2013).
\bibitem{MU3E} A. Blondel, {\it{et al.}} ({\mue[3]} Collaboration), 
  arXiv:1301.6113 (2013).
\bibitem{ref:HVMAPS}    I.~Peric,
  %``A novel monolithic pixelated particle detector implemented in high-voltage CMOS technology,''
  Nucl.\ Instrum.\ Meth.\ A582, 876 (2007).
%  doi:10.1016/j.nima.2007.07.115
  %%CITATION = doi:10.1016/j.nima.2007.07.115;%%
  %109 citations counted in INSPIRE as of 11 Nov 2018
\bibitem{Papa2018} A. Papa, NuFact 2018, Blacksburg, Virginia USA, \\
  {\tt https://indico.phys.vt.edu/event/34/contributions/701}
\bibitem{MELC} R.M. Dzhilkibaev and V.M. Lobashev, Sov.J.Nucl.Phys. 49(2), 
  (1989) 384.
\bibitem{MUSIC} S. Cook, {\it{et al.}}, Phys. Rev. Accel.  Beams 20(3), (2017) 030101.
\bibitem{COMET-I-TDR} R. Abramishvili, {\it{et al.}} 
  ({\comet} Collaboration), \\
  {\tt http://comet.kek.jp/Documents\_files/PAC-TDR-2016/COMET-TDR-2016\_v2.pdf}
\bibitem{MU2E} L. Bartoszek, {\it{et al.}} ({\mue[2]} Collaboration),  
  arXiv:1501.05241 (2015).
\bibitem{Mu2e-tracker} M. Lee (on behalf of \mue[2] Collaboration), 
  Nucl. Part. Phys. Proc., 273 (2016) 2530.
\bibitem{Mu2e-calorimeter} N. Atanov, {\it{et al.}}, 
  IEEE Trans. Nucl. Sci., Vol 65, N. 8, (2018) 2073.
\bibitem{COMET-ESPP} COMET submission to the European Strategy for 
  Particle Physics 2020, and references therein.
\bibitem{Mu2e-II-EOI} F. Abusalma, {\it{et al.}} ({\mueii} Experiment),
  arXiv:1802.02599 (2018).

\end{thebibliography}
\end{document}